\documentclass[aps,preprint,onecolumn]{revtex4}
\usepackage{dcolumn}
\usepackage{bm}
\usepackage{amssymb}
\usepackage{amsmath}
\usepackage{amsfonts}
\usepackage[]{graphicx}

\begin{document}

\title{The PBR theorem seen from the eyes of a Bohmian}

\author{Aur\'{e}lien Drezet}
\affiliation{Institut N\'eel UPR 2940, CNRS-University Joseph
Fourier, 25 rue des Martyrs, 38000 Grenoble, France}

\begin{abstract}
The aim of this paper is to present an analysis of the new theorem
by Pusey, Barrett and Rudolph (PBR) concerning ontic and epistemic
hidden variables in quantum mechanics~\cite{PBR,Leifer1}. This is a
kind of review and defense of my previous critical analysis done in
the context of Bohmian mechanics. This is also the occasion for me
to review some of the fundamental aspects of Bohmian theory rarely
discussed in the literature.
\end{abstract}
\maketitle
\section{A not too `Bohring' Introduction to Bohm (I hope)}
I am a Bohmian (i.e. a `de Broglian') which means somebody believing
in the pertinence of the pilot wave theory proposed by de Broglie in
1926-27 and rediscovered by Rosen in 1945 and Bohm in 1952 (see the
book by Holland~\cite{Holland}). What is pilot wave theory? A
completely deterministic and neat approach at the fundamental level
involving trajectories and dynamical laws for point-like quanta  (at
least in its original version). This quantum interpretation which
contrasts with the one proposed by Bohr Heisenberg and others is
done in such a way as to agree completely with quantum mechanics
rules and in particular is tuned to reproduce every statistical
prediction given by the usual formalism (this is why we speak about
an interpretation of the quantum formalism). The theory works not
only for a single particle, but also for systems of several
entangled objects (even though entanglement was not clearly defined
in 1927) such as particle beams or molecules. Furthermore, the
theory is completely nonlocal in the sense defined by Bell with his
famous theorem of 1965. Therefore, the theory although deterministic
is able to describe subtle quantum effects such as correlations
(i.e., the EPR paradox) and interferences (i.e., the wave particle
duality) and provides a clear ontology for understanding the quantum
world by solving all the measurement paradoxes. The reaction to this
proposition was from the beginning very emotional and the theory of
de Broglie and Bohm was often named `metaphysical' or `ideological
superstructure' and even recently accused of being `surrealistic'
(see for example refs.~\cite{Scully,Vaidman}). The main reasons for
the strong opposition is that pilot wave says that things which are
not experimentally determinable are however determined in a very
precise way by dynamical laws (the so called guidance equations of
de Broglie). But, since the pilot wave agrees with quantum mechanics
it should also certainly accept the Heisenberg uncertainty and the
results concerning wave particle duality with the double-hole
experiment. How could that be? Indeed, pilot wave agrees with all
that but in a very peculiar way. To understand that, I remind you
briefly what is the point of view of Bohr and Heisenberg on this
topic. The argumentation focuses on the famous double-hole
interference experiment done with single electron or photon and
which shows that a particle could be influenced by the hole through
which it is not going to pass in order to create an interference
pattern. This is a kind of paradox if we try to think in term of a
particle path going from only one hole and which `obviously' should
not care about the `remote' presence of the second hole. For Bohr
and Heisenberg this paradox should be removed. `Fortunately', they
wrote, the presence of the `particle', i.e., the `trajectory' can
not be detected at both holes without disturbing the fringes.
Therefore, at least at the experimental level, no contradiction like
\emph{to be at A and not at A at the same time} can occur.  Bohr and
Heisenberg emphasize that the result is actually worst that a naive
picture of the uncertainty principle could \emph{apriori} let us to
believe. Indeed, this naive semi-classical picture would say that
the measurement always disturbs but that `OK we could still may-be
preserve, at least conceptually, trajectories even if they are
hidden'. However, quantum mechanics predicts  that even a very small
interaction which localize the particle, say in only one arm of an
interferometer but not in the other (the spatial precision is not so
huge here since the interferometer can be very big), will disturb
and destroy the subsequent fringes. Therefore, it seems that
hypothetical trajectories have no meaning in the experimental world,
and since they can not be investigated they are metaphysical.
Quantum mechanics textbooks are full of examples like the previous
one discussed either in term of momentum `kicks' \emph{\`{a} la}
Heisenberg or Feynman or involving more sophisticated devices and
entanglement machineries. All the practitioners of the orthodox
school generally emphasize that there is \emph{no other choice}: in
the quantum world we have to abandon our habits our clean logics and
accept that things can not be fully described by the classical
categories such as position $\mathbf{X}(t)$ and velocity
$\dot{\mathbf{X}}(t)$ characterizing locally the system and evolving
deterministically with time. Following Bohr and his complementarity
principle one must choose which variable we want to experimentally
define and we can then unambiguously calculate the probability of
occurrence for such events using the quantum rules. However, these
experimental contexts sometimes exclude each other (i.e., they are
complementary like for example experimental arrangements for
measuring either $x$ and $p$ for \emph{a same} particle) and we must
definitely renounce to our classical illusions such as trajectories
and paths existing independently of the observation. Of course, the
time evolution $\mathbf{X}(t)$ disappears completely from the
discussion  and we are allowed only to speak about the probability
$dP(\mathbf{X},t)$ to observe the system with the value $\mathbf{X}$
at the time $t$. If we don't measure $\mathbf{X}$  then it has no
actualized reality; it was only a potentiality at the given time
$t$.   The subsequent evolution of the then undisturbed
wave-function  $|\Psi(t')\rangle $ will give other potentialities at
a future time  $t'$ which again will or not be actualized in our
experimental world depending on your will to
measure it or not.\\
\indent If experimentally you can not determine a trajectory with a
too large precision, i.e., at least nor large enough to observe both
the path and fringes with \emph{a same} particle, what could be the
interest of such a pilot wave dynamics? This is a clear drawback of
the Bohmian approach and it explains why it was so attacked strongly
by Heisenberg, Pauli and many others. Although pilot wave solves in
a neat way the measurement problem by fixing an ontology it also
brings us parameters  which somehow stay `hidden' and therefore
apparently metaphysical. However, I think this reaction exaggerated.
First, we could remark that Heisenberg and Bohr are not completely
fair concerning trajectories when they say that these paths have no
existence. Actually, they go too far since their claim can not be
proven either and are even contradicted by the pilot wave mere
existence (as it was emphasized by de Broglie and Bohm in 1951-52).
In particular, it is important to remind that von Neumann
demonstrated in the 1930's a famous theorem forbidding the existence
of such a kind of hidden variable model and until the 1980's it was
often quoted as a final impossibility proof for the existence of
trajectories, even though pilot wave was already a counter example,
and even after Grete Hermann and later John Bell showed that the
axiomatic of the theorem is not general enough to get to the von
Neumann expected theorem. I think that the Copenhagen interpretation
should be amended seriously at least on that point by replacing the
world \emph{non existent} by something like \emph{experimentally
hidden without breaking the fringe coherence}. But is this really
true? Are particle paths completely hidden at the experimental
level? This is not actually totally the case. In recent years much
more was written on weak values as defined by Aharonov, Albert and
Vaidman~\cite{Albert} and in particular on the possibility to
identify a certain weak value $\mathbf{A}_w$ with the velocity field
$\mathbf{\dot{X}}(t)$ attributed  precisely by the pilot wave to the
particle located at $\mathbf{X}(t)$. Actually, this was
experimentally demonstrated \cite{Kocsis} showing that the Bohmian
trajectories can have an experimental reality.  There is however no
contradiction with what was find and discussed before. The trick is
indeed to realize that a weak measurement is not  done on a single
individual unlike the strong projective measurement. Weak
measurement is weak and requires a large population of particles to
get the trajectories. Therefore, in all these examples the
Heisenberg principle stays valid: we can not detect fringes and path
for \emph{a same} particle. Therefore, the sentence
\emph{experimentally hidden} means in reality \emph{experimentally
hidden at the single particle level}. But, I would like to point out
that even this apparently prudent analysis is not exempt of critics.
Indeed, beside the weak measurement protocol  Aharonov and Vaidman
also defined what they called a protective measurement
protocol~\cite{Vaidman2}. This is a very interesting method
focussing   on the fact that  in some conditions we can define a
system  $S$ evolving very slowly and gently (i.e. adiabatically)
which can be coupled to a meter which evolves very strongly into a
well distinguishable state. The result of the protocol will not give
us a way to record precisely the spectrum of an observable $A$ of
the system (i.e. unlike in a von Neumann protocol) but either, will
give us the new possibility to measure its average value
$\langle\psi_ S|A|\psi_ S\rangle$. This is very interesting in the
context of pilot wave for several reasons. First, since $\langle
A\rangle$ can be for example the probability density
$\rho(\mathbf{X}_0)=\langle\psi_
S|\mathbf{X}_0\rangle\langle\mathbf{X}_0|\psi_ S\rangle$ or the
current $J(\mathbf{X}_0)=\langle\psi_
S[|\mathbf{X}_0\rangle\langle\mathbf{X}_0|\mathbf{P}-\mathbf{P}|\mathbf{X}_0\rangle\langle\mathbf{X}_0|]/(2mi)|\psi_
S\rangle$ of the particle (with mass $m$) at point $\mathbf{X}_0$,
one could at first argue (like in ref.~\cite{Aharonov}) that the
protocol proves once again the surrealistic nature of the Bohmian
trajectories. Indeed, the protective measurement protocol can be
used to `detect' the particle at points where the Bohmian particle
never approaches. This reasoning is based on the fact that for a
real wave function $\langle \mathbf{X}|\psi_ S\rangle$ the Bohmian
particle is not moving at all (i.e., $\dot{\mathbf{X}}(t)=0$) so
that even if the particle is fixed at position
$\mathbf{X}_1\neq\mathbf{X}_0$ the protective measurement will allow
to measure $\rho(\mathbf{X}_0)$. How could that be?  Although I will
not here answer to that in details I can provide a simple
qualitative explanation: particle is not everything in the pilot
wave. For a Bohmian the wave is also a fundamental ingredient so
that the force exerted on a particle depends not only on the
`contact' potential proposed in ref.~\cite{Aharonov}  but also on a
quantum potential which can acts in some non classical but
completely deterministic way. This is enough to justify how the
dynamics of the pointer is affected in some nonlocal way by the
quantum interaction. I actually developed a complete Bohmian
reasoning in \cite{Drezet} as a reply to ref.~\cite{Aharonov}, see
also the forthcoming chapter in the Book `Protective Measurement and
Quantum Reality' edited by Shan Gao~\cite{Gao}). There is however an
other reason why protective measurement is interesting in the
context of Bohmian mechanics. Although I didn't emphasized that
point enough in the past this is actually much more important.
Indeed, protective measurement is done at the single particle level
which means that even a single pointer measurement allows us to
determine $\rho(\mathbf{X}_0)$ or $J(\mathbf{X}_0)$. But since the
operators associated with $\rho(\mathbf{X}_0)$ or $J(\mathbf{X}_0)$
commute actually nothing forbid us to measure $\rho(\mathbf{X}_0)$
and $J(\mathbf{X}_0)$ together (for example with two pointers).  But
now, for a Bohmian this is a bit of magic because we have a way to
measure at the single particle level the ratio
$J(\mathbf{X}_0)/\rho(\mathbf{X}_0)$ which is nothing else that the
particle velocity. It is thus not anymore justified to say that the
Bohmian  velocity is not an observable. Of course  in someway the
Heisenberg uncertainty principle is not in question since the
protective measurement is not a projective detection of the particle
position at $\mathbf{X}_0$. We don't have access to the actual
trajectory followed by the particle because knowing the velocity is
not enough:  we should also have the actual position but this would
require a projective method. However, we could imagine the following
operations: first make a protective measurement to obtain the
velocity at  $\mathbf{X}_0$, then measure projectively for the same
particle its position  $\mathbf{X}$. Subsequently, retain only those
cases where the projective measurement gives
$\mathbf{X}=\mathbf{X}_0$. We have thus both the particle and
velocity for the same particle at the same time! Note that the
future evolution will be however random since the projective
measurement is very intrusive. Still, this result is  I think
remarkable.  I point out that it relies on the definition of the
time scales involved in the process. Indeed, if by protective we
mean adiabatic and very slow then the complete two-measurements
procedure proposed here will have only meaning if the Bohmian
velocity is very small so that it will still makes perfectly sense
to speak about a velocity and position
recorded at the same time for one particle.\\
\indent There are other reasons for defending Bohmian mechanics. One
of them is that it provides finally a kind of intelligibility which
is absent from the Copenhagen interpretation. Indeed, since for Bohr
we can not say anything about the system between measurements, it
means, like it was shown by Wigner, that an observer can stays in a
ubiquitous quantum state without clean ontological status before a
second observer finalizes his experiment. How could that be and what
does it mean? If we speak only about epistemic there is no real
problem since knowledge is indeed relative. However, if we speak
about ontology this is a non sense (this is also the main message of
the Schrodinger cat paradox I think). But if we follow Heisenberg
and his quantum/classical `cut' this conclusion is unavoidable.
Ultimately, the Universe as a whole becomes an issue. Does god
existence (with a Ph.D) proven to be necessary for collapsing the
wave function of the Universe? This seems extremely difficult to
believe for me. This is an example of twilight zone which surrounds
Bohr-Heisenberg interpretation and this the reason why for me
Bohmian is superior to Copenhagen. Still, one could perhaps
criticize Bohmian mechanics on a different level. I remind indeed
that for a non relativistic particle of mass  $m$ the pilot wave
particle velocity is given by the de Broglie guidance formula
\begin{eqnarray}
\frac{d}{dt}\mathbf{X}(t)=\frac{\hbar}{2mi}\Psi(\mathbf{X},t)^\ast\stackrel{\leftrightarrow}{\boldsymbol{\nabla}}\Psi(\mathbf{X},t)/|\Psi(\mathbf{X},t)|^2=\frac{\mathbf{J}(\mathbf{X},t)}{|\Psi(\mathbf{X},t)|^2}
\end{eqnarray}
where $\mathbf{J}$ is the Madelung probability current arising from
Schr\"{o}dinger equation. However, from local conservation we have
$\boldsymbol{\nabla}\cdot\mathbf{J}(\mathbf{X},t)+\partial_t|\Psi(\mathbf{X},t)|^2$.
It is thus clear that we can add a rotational
$\boldsymbol{\nabla}\times \mathbf{C}(\mathbf{X},t)$ to the current
without changing the conservation. How could we be sure that our
velocity formula is the good one? Pilot wave can not answer that
univocally without calling to an other principle. For example one
could try to invoke some Galilean or Lorentzian symmetries or
principles~\cite{Durr}. We could also invoke  weak measurement or
protective measurement for giving an empirical support to some
Bohmian concept not anymore so hidden.  The answer to be given for
this lack of univocity is not clear but for me it actually means
that Bohmian mechanics is only a temporary expedient waiting for
something of better, i.e., for a theory in which the pilot wave
dynamics will appear as a consequence more than a postulate. An
other element leads to the same conclusion: the wave acts on the
particle but the reciprocal is not true. Therefore, it seems that
the Bohmian quantum force is only an effective trick and that
something of deeper is hidden here waiting for further
investigations and discoveries (may be along the path proposed by de
Broglie with its double solution program). I also mention a
difficulty with the energy concept: For a general quantum state the
actual Bohmian Energy defined by $E=-\partial_t S(\mathbf{X},t)$,
where $S/\hbar$ is the wave function local phase, is not in general
a constant even in the absence of any external potential. It is for
me very difficult to accept such a feature for a final theory: the
total energy should be a constant in the absence of external forces.
Probably the energy definition is not so good here.  This again,
motivates for further investigations beyond the pilot wave. In the
same vain, sometimes the Bohmians speak about `empty waves'
\cite{Hardy} when for example a wave pack splits into several
branches and when a particle chooses only one. The others branches
are clearly empty of particles but are the waves still there in the
branches? If the quantum potential has a reality independent of the
particle the answer is `yes! certainly ' but there is no proof of
that and empty waves have not been directly detected yet. Once
again, I think these are strong arguments for going beyond
the pilot wave approach  and that quantum mechanics will be superseded by something else (this was the conviction of de Broglie by the way).\\
\indent This is a long introduction to justify my quantum
realist/determinist position.  But it serves only as a motivation
for the next short section where I will describe the PBR theorem and
its relation with Bohmian mechanics. PBR is an important result
obtained at the end of 2011 by Pusey Barret and Rudolph concerning
the relation of epistemic and ontic in hidden variable theories.  In
the long tradition started with Bell (or more honestly von Neumann)
its aims is to give experimental bounds to the allowed models that
quantum realists can propose. Bell, focussed on non-locality,  a
feature of Bohmian mechanics, and PBR were interested by the
experimental definition of epistemic models.  I will shortly review
the PBR result~\cite{PBR} (without the demonstration) and explains
why pilot wave escapes the conclusions. Still the theorem is true if
we add an axiom. I actually found this rather simple result already
in 2011 immediately after that the preprint of PBR circulated on the
web but the work was published only later for editorial reasons. I
also discussed this subject with M. Leifer on his blog page early in
2012~\cite{Leifer1} (but we disagreed on the conclusion as it is
also shown in his recent manuscript~\cite{Leifer}: the current paper
is also a kind of reply to him). For more details on the proof the
interested readers could find some of my earlier manuscripts on
Arxiv (see refs.\cite{Drezet2,Drezet3}) and compare with a
independent work by M.Schlosshauer and A. Fine~\cite{Fine} who
clearly discovered the same result independently and simultaneously.
\section{the PBR result and its meaning for a Bohmian}
\indent What is PBR theorem? the demonstration that epistemic models
are forbidden in quantum mechanics.  Why epistemic models? Epistemic
or knowledge interpretations have a long tradition in quantum
mechanics. Einstein was a strong  defender of such approaches and
for him it meant that quantum mechanics was a kind of statistical
mechanics like in the classical world but waiting for something of
better with a clean deterministic foundation (again like classical
mechanics).  For  Einstein, quantum mechanics was a bit like
thermodynamics before the works of Clausius, Maxwell and Boltzmann
on statistical physics. Actually, this is not really different from
the de Broglie and Bohm point of view and we should not forget that
Einstein proposed already in 1907 that particle of light should be
envisioned as a kind of singularity riding atop a guiding
electromagnetic field (this is the de double solution program of de
Broglie). De Broglie succeeded where Einstein failed and the pilot
wave of de Broglie-Bohm indeed justifies the existence of
probability by a statistical mechanical argument like Boltzmann or
Gibbs did with Newton laws.  By Epistemic models PBR meant actually
a sub-class of this kind of statistical model but they didn't
realize it in their paper. Before to come to this let go to the
first step of the PBR theorem which is purely quantum in the sense
of the formalism. In the simplest version PBR considered two non
orthogonal pure quantum states $|\Psi_1\rangle=|0\rangle$ and
$|\Psi_2\rangle=[|0\rangle+|1\rangle]/\sqrt{2}$ belonging to a
2-dimensional Hilbert space $\mathbb{E}$ with basis vectors
$\{|0\rangle,|1\rangle\}$. We will limit ourself to this example for
the discussion since the details are not so important here. Using a
specific measurement protocol $M$ with basis $|\xi_i\rangle$
($i\in[1,2,3,4]$) in $\mathbb{E}\otimes\mathbb{E}$ which precise
form is here irrelevant (see ref.\cite{PBR}) PBR deduced that
$\langle\xi_1|\Psi_1\otimes\Psi_1\rangle=\langle\xi_2|\Psi_1\otimes\Psi_2\rangle=\langle\xi_3|\Psi_2\otimes\Psi_1\rangle=\langle\xi_4|\Psi_2\otimes\Psi_2\rangle=0$.
which means that some probabilities cancel with such protocols.  Now
in order to see the contradiction  we go to the second step and try
to introduce an hypothetical hidden variable model reproducing the
statistical features of quantum mechanics.  This is clearly the
classical methodology proposed by Bell.  Bell introduced  `hidden
variables' $\lambda$ which in the Bohmian language could be the
possible coordinates of the particles at the initial time.  Here, I
will be more precise that PBR because I want to emphasize later some
limitations on the reasoning. First, consider a quantum state
$|\Psi\rangle$ and an observable $A$ with eigenvalue $\alpha$. The
probability of occurrence for $\alpha$ will be given by
\begin{equation}
|\langle\alpha|\Psi\rangle|^2=P(\alpha,\mathbf{a}|\Psi)=\int
P(\alpha,\mathbf{a}|\lambda)\rho(\lambda|\Psi)d\lambda.
\end{equation}
In this notation  we introduced  the hidden variable distribution
$\rho(\lambda|\Psi)$ and the conditional probability
$P(\alpha,\mathbf{a}|\lambda)$ (such as $\sum_{\alpha}
P(\alpha,\mathbf{a}|\lambda)=1$ by definition of a conditional
probability) defining the `likehood' for the system to evolve from
its initial state (characterized by its hidden variable $\lambda$,
and its wave function) to a state where the eigenvalue $\alpha$ will
be actualized (i.e. after a projective measurement characterized by
some external parameters $\mathbf{a}$ such as the spin analyzer
direction in a Stern Gerlach experiment). These definitions are very
classical-like since the dynamic or `ontic' state  should be
decoupled from its epistemic counterpart in agreement with the
Boltzmann-Gibbs statistical approach. Of course,
$\rho(\lambda|\Psi)$ is supposed  to be independent of $\mathbf{a}$
since causality is expected to hold from past to future and if your
reject retro-causal, some `magical' conspiracy or super
deterministic approaches \`{a} la Costa de Beauregard or John Cramer
(e.g., the very interesting transactional interpretation). Now, in
the PBR reasoning we should write
\begin{eqnarray}
|\langle\xi_i|\Psi_j\otimes\Psi_k \rangle|^2=\int
P_M(\xi_i|\lambda,\lambda')\varrho_j(\lambda)\varrho_k(\lambda')d\lambda
d\lambda'
\end{eqnarray}
where $i\in[1,2,3,4]$ and $j,k\in[1,2]$.  Actually, in their paper
PBR didn't use such notations but these obviously simplify the
reasoning like they did for Bell. In this PBR model there is an
independence criterion at the preparation since we write
$\varrho_{j,k}(\lambda,\lambda')=\varrho_j(\lambda)\varrho_k(\lambda')$.
This is a very natural axiom and for example such an axiom would be
justified in the Bohmian interpretation where the hidden parameters
are the initial coordinates $\mathbf{X}_1(0)$ and $\mathbf{X}_2(0)$
of the particles in the incident wave-packets (Although we are here
speaking about Q-bit this is not a problem: Bohm works also for
spins but here the Q-bits could simply belong to a sub-manifold of
the full hilbert space like it is for instance with two-energy-level
systems. Therefore, spatial coordinates are still relevant). In
these equations we again introduced the conditional `transition'
probabilities $P_M(\xi_i|\lambda,\lambda')$ for the outcomes $\xi_i$
supposing the hidden state $\lambda,\lambda'$ associated with the
two independent Q-bits are given. The fundamental point here is that
$P_M(\xi_i|\lambda,\lambda')$ is independent of $\Psi_{1},\Psi_{2}$.
Obviously, we should have
$\sum_{i=1}^{i=4}P_M(\xi_i|\lambda,\lambda')=1$. It is then easy
using all these definitions and conditions to demonstrate that we
must necessarily have
\begin{eqnarray}\varrho_2(\lambda)\cdot\varrho_1(\lambda)=0 &
\forall \lambda,\end{eqnarray} i.e., that $\varrho_1$ and
$\varrho_2$ have nonintersecting supports in the $\lambda$-space.
This constitutes the PBR theorem for the particular case of
independent prepared states $\Psi_1,\Psi_2$ defined before (but PBR
generalized their results for more arbitrary states using similar
and astute procedures described in ref.~\cite{PBR}). What are the
implications of such a result?  If we identify the conditions
imposed by PBR on the hidden variable models with what should be
naturally expected from any ontological model having a statistical
ingredient, then we could conclude that such models are nor really
statistical. Indeed, from Eq.~4 we deduce that the density of
probabilities $\varrho_{\Psi_1}(\lambda)$
$\varrho_{\Psi_2}(\lambda)$ for any two quantum states $\Psi_1$ and
$\Psi_2$ are necessarily \emph{not} overlapping in the
$\lambda-$(phase) space. Therefore, it will be like if we have
necessarily a delta distributions $\delta^3(q-X(t))\delta^3(p-P(t))$
in classical mechanics. This kind of model could hardly be called
statistical at all? If this theorem is true (and mathematically it
is) then it would apparently make hidden variables completely
redundant since it would be always possible to define a relation of
equivalence between the $\lambda$ space and the Hilbert space:
(loosely speaking, we could in principle make the correspondence
$\lambda\Leftrightarrow\psi$). In other words, it would be as if
$\lambda$ is nothing but a new name for $\Psi$ itself!\\
\indent However the PBR reasoning doesn't fit with the Bohmian
mechanics framework and therefore it is not difficult to see that
the reasoning obtained by PBR can not hold for such a theory. First,
observe that  for pilot wave we have  both $\mathbf{X}(t)$  and
$\Psi(\mathbf{X},t)$ as ontological variables  and since Born's rule
occurs then by definition
$\rho_\Psi(\mathbf{X},t)=|\Psi(\mathbf{X},t)|^2$ defines  in the
pilot wave model the probability of presence  for the particle. If
we consider the initial state at the initial time $t_0$ we have
$\rho_\Psi(\lambda):=|\Psi(\mathbf{X},t_0)|^2$. This is an epistemic
distribution of hidden variables guided by the wavefunction
$\Psi(\mathbf{X},t)$. Clearly, for two given states $\Psi_1$ and
$\Psi_2$ (orthogonal or not) we have in general
$\rho_{\Psi_1}(\lambda)\cdot\rho_{\Psi_2}(\lambda)\neq0$ in
contradiction with Eq.~4 and PBR statement. To see why it is like
that we first point out that Bohm model is deterministic. Therefore,
for a given $\lambda_0:=\mathbf{X}(t_0)$ we know that the evolution
of the system in a projective measurement will also be
deterministic. After the measurement is done the particle is
actually in one of the allowed eigenvalues $\alpha_0$ (supposed
discrete here for simplicity) and we can write
$\alpha_0=A(\lambda_0,\mathbf{a},\Psi_0)$. We should consequently
write Eq.~2 with
\begin{eqnarray}
P\left(\alpha|\mathbf{a},\lambda,\Psi_0\right)=\delta_{\alpha,A(\lambda,\mathbf{a},\Psi_0)}=
0 \textrm{ or } 1
\end{eqnarray}
where $\delta$ is the Kronecker symbol, since for one given
$\lambda$ only one trajectory is allowed (this model of course
satisfies trivially the condition $\sum_\alpha
P(\alpha|\mathbf{a},\lambda,\Psi_0)=1$). Equivalently, the actual
value $A(\lambda,\mathbf{a},\Psi_0)=\sum_{\alpha}\alpha
P\left(\alpha|\mathbf{a},\lambda,\Psi_0\right)$ can only takes one
of the allowed eigenvalues $\alpha$ associated with the hermitian
operator $A$.  Such kind of notations were used by Holland in his
book \cite{Holland} (see also~\cite{vigoureux}). What is fundamental
here is that Eq.~5 depends on $\Psi_0$(the initial wave function) in
a explicit way. Still, beside this contextually the Bohm model is a
clean statistical model and there is no reason which can forbid us
to call it an epistemic model. This discussion shows however that
pilot wave is not a banal classical model it contains a wave
function $\Psi_0$ which have a particular status: it guides the
particle and at the same time it characterizes completely the
statistical ensemble for a given protocol. While, $\lambda$ can
fluctuate in the ensemble (corresponding to the different possible
values for $\mathbf{X}(t_0)$) $\Psi_0$ is instead a kind of
dynamical constraint belonging to an ensemble like was the action or
the energy in the old Hamilton-Jacobi theory: $\Psi$ guides the
particles and characterize the statistical ensemble~\cite{footnote}.
Moreover, Eq.~2 is now modified and we should write
\begin{equation}
|\langle\alpha|\Psi\rangle|^2=\int
\delta_{\alpha,A(\lambda,\mathbf{a},\Psi_0)}\rho(\lambda|\Psi)d\lambda.
\end{equation} to take into account Eq.~5.
Clearly, this means that PBR Eq.~3 should be modified as well to
include this new contextual feature:
\begin{eqnarray}
|\langle\xi_i|\Psi_j\otimes\Psi_k \rangle|^2=\int
P_M(\xi_i|\lambda,\lambda',\Psi_j,\Psi_k
)\varrho_j(\lambda)\varrho_k(\lambda')d\lambda d\lambda'.
\end{eqnarray}
However, now we have lost the secret ingredient allowing us to
obtain Eq.~4 which implies that the PBR derivation doesn't hold
anymore! (details are discussed elsewhere \cite{Drezet2,Drezet3}.
Part of the language used here was also introduced long ago by Fine
~\cite{Fine2} and discussed by me in a different context
~\cite{vigoureux,Drezet4}). What does it mean? The ontic-epistemic
framework used by PBR suggested that there is a clean separation
between ontic and epistemic approaches. This is motivated by the PBR
sentence `The statistical view of the quantum state is that it
merely encodes an experimenter's information about the property of a
system. We will describe a particular measurement and show that the
quantum predictions for this measurement are incompatible  with this
view' \cite{PBR}. By `merely'  PBR meant certainly something like
classical statistical mechanics but what about Bohmian theory? Are
they really ontic for them? Does PBR simply ignore it? I found that
suspicious since Harrrigan-Spekkens start their
paper~\cite{speckens} (cited in \cite{PBR}) by the following
definition: `We call a hidden variable model $\psi-$ontic if every
complete physical or ontic state in the theory is consistent with
only one pure quantum state; we call it $\psi-$ epistemic if there
exist ontic states that are consistent with more than one pure
quantum state'. Now, as explained, Bohmians proposed since 1927
statistical interpretations where the wavefunction plays a dual
role. $\Psi$ guides the particles but also justify the quantum
statistical observations with some clear epistemic elements. Clearly
, for a Bohmian the wavefunction is definitely \emph{not only} a
simple label to our epistemic knowledge but it is any way
\emph{also} such a label! In agreement with the previous quotation I
would thus say that pilot wave is in part also epistemic but this is
not actually the case in the ontological framework of these authors.
They actually classified Bohmian mechanics as `$\psi$-supplemented'
(a sub class of `$\psi$-ontic') meaning that additionally to
$\Psi(\mathbf{x},t)$ we must add some hidden supplementary variables
$\mathbf{X}(t)$. Somehow, I could agree also with this second
definition which seems however to contradict my previous choice. So
what! Is Bohmian mechanics epistemic or ontic? This is very
confusing (i.e., not only for me; see for example
Feintzeig~\cite{Feintzeig} who is also clearly disturbed by that).
Since, the paper~\cite{speckens} played an important role in the
work of PBR I think that there is a kind of language ambiguity in
the reasoning. May be, PBR could reply to the critics by saying like
Leifer (in his analysis of the work by me and M. Schlosshauer and A.
Fine:  ref. \cite{Leifer} pages 60-63): `if your conditional
probabilities for measurement outcomes depend on the wavefunction
then the wave function is ontic and there is nothing left to prove.'
I indeed received few emails along that direction. However, for me
the central point is not that the wave function is ontic (I have no
doubt about that: see the first sentence of this article), but that
epistemic is not orthogonal to ontic and that therefore the wave
function is also an epistemic carrier. Interestingly, Leifer agrees
in the same paper that `the scope of the PBR theorem is restricted
to the case where this conditional independence holds'. However he
then adds: ` but this is part of the definition of the term ``ontic
state'', rather than something than can be eliminated in order to
arrive at a more general notion of what it means for  a model to be
$\psi-$epistemic that still conveys the same meaning'. In other
words he recognizes that the PBR derivation doesn't hold if you
reject the $\Psi-$independence in the conditional probabilities but
that I modified the definition of epistemic used by PBR. Clearly, we
don't have the same definition of what is to be ontic and epistemic.
For me Bohmian mechanics is both ontic \emph{and} epistemic while
for Leifer and some others it is purely ontic. This looks like a old
problem of semantic. Semantic plays indeed a role in this debate.
PBR, Leifer and others call ontic respectively  what M. Schlosshauer
and A. Fine~\cite{Fine} called 'segregated models' and `mixed
models'. I clearly prefer the vocabulary of M. Schlosshauer and A.
Fine although personally I would simply use something like non
overlapping and overlapping distributions instead of segregated and
mixed (this would agree with the figure 1 of the PBR paper
\cite{PBR}, see also~\cite{Leifer1}). Also, I completely agree with
them~\cite{Fine} when they wrote: `we find this terminology less
charged than the terms ``$\psi$-epistemic'' and ``$\psi$-ontic''
that PBR adopt from~\cite{speckens}[my reference]'. In particular,
epistemic is very much charged in the context of probability theory
where the objective or subjective nature of the concept is often
debated. Furthermore, in classical mechanics even a simple
trajectory is a solution of Liouville equation and corresponds to an
`epistemic' density of probability
$\rho(q,p,t)=\delta^{3N}(q-X(t))\delta^{3N}(p-P(t))$ associated with
a perfect knowledge.  For the word ontic the situation is even
worst. Ontic, is a philosophical word and its definition is a bit
like God: everyone knows what it means but nobody agrees...I suggest
that the use of such a charged vocabulary is responsible for the
confusion surrounding this PBR theorem, therefore semantic is indeed
here a problem. In the same vain, I would like to precise that I
first learned about the PBR theorem version mainly through the Arxiv
2011 preprint of the PBR manuscript (compare with the final
manuscript \cite{PBR}) and from the early pedagogical presentation
by Leifer~\cite{Leifer}, and Barrett (done at Oxford the 12$^{th}$
of March 2012~\cite{Barrett}). In all these works, the authors
clearly consider the opposition ontic-epistemic in the sense
segregated-mixed which is unambiguous. However, nowhere the
postulate that $P(\alpha|\mathbf{a},\lambda)$ should be independent
of $\Psi_0$ is even mentioned. This is the reason why I can fairly
conclude that they didn't included this axiom in their reasoning.
For example Barrett mentions at slide 15 of his presentation that
$P(\alpha|\mathbf{a},\lambda)$ is a natural axiom of Bell whereas
Bell never postulated such a constraint. Furthermore, at slides
16-17 the opposition ontic$\Leftrightarrow$epistemic is done in such
a way as to oppose the non-overlapping$\Leftrightarrow$overlapping
distribution like if every thing was there. But, since the missing
postulate $P(\alpha|\mathbf{a},\lambda,\Psi_0)\Leftrightarrow
P(\alpha|\mathbf{a},\lambda)$ is not mentioned it seems to play no
role at all in the reasoning (this is not surprising since it
doesn't appear either in the Harrigan-Spekkens
paper~\cite{speckens}). However, once again, the opposition
$P(\alpha|\mathbf{a},\lambda,\Psi_0)\Leftrightarrow
P(\alpha|\mathbf{a},\lambda)$ has a clear ontological and epistemic
status (as important that the one associated with the overlapping or
non overlapping density of states) and it \emph{must not} be
neglected otherwise the theorem is simply incomplete. We can also
better appreciate this point by comparing ~\cite{PBR,Leifer,Barrett}
with refs.~\cite{Fine,Fine2,Drezet4} where a clear discussion of
what it means to include $\Psi_0$ in the probability
$P(\alpha|\mathbf{a},\lambda,\Psi_0)$ is done.\\
\indent My critical analysis of the PBR theorem was however not
intended to be semantical. It was not done for rejecting the
complete PBR reasoning but only to show that the presentation of the
theorem should be amended in order to make it general. The postulate
that $P(\alpha|\mathbf{a},\lambda,\Psi_0)$ should be independent of
$\Psi_0$ is a critical part of the PBR derivation and should be
explicitly included in order to see the limitations of the theorem
and re-enforces his strength. Let me propose a version of the PBR
theorem:\\
\indent \emph{\underline{PBR theorem (amended version):}}\\
\emph{If $P\left(\alpha|\mathbf{a},\lambda,\Psi_0\right)$ is
independent of $\Psi_0$ then Eq.~4 holds necessarily. In the
opposite case this is not necessarily true.}\\
The inclusion of this additional postulate concerning conditional
probabilities has important consequences since it will shed some
light on the
properties of Bohmian mechanics (a bit like Bell's theorem did).\\
\indent Consider, for example the simple beam-splitter experiment
shown on Figure 1. If we send a  single photon state
$|\Psi_1\rangle$ through the input  gate 1. The wave packet splits
and we will finish with  a probability $P(3|1)=1/2$ to detect the
photon in the exit 3
\begin{figure}[h]
\begin{center}
\begin{tabular}{c}
\includegraphics[width=8.5cm]{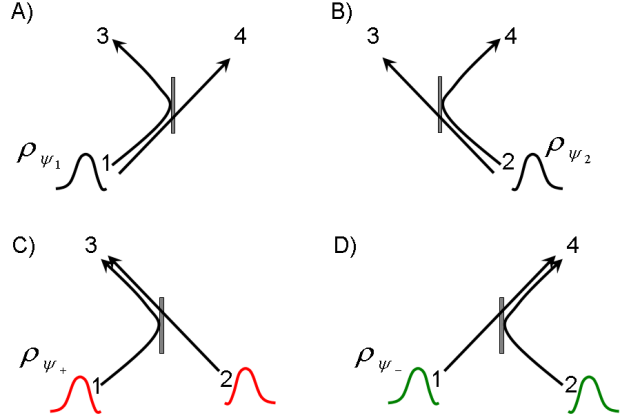}
\end{tabular}
\end{center}
\caption{An example showing that the `old' axiomatic of PBR can not
be applied to a Bohmian like model. Cases A and B correspond to wave
packets impinging from one of the 2 beam-splitter entrances.  Exits
3 and 4 are both allowed.   In case C and D where a superposition of
wave packets  interfere coherently  the exits 4 and respectively 3
are forbidden. In Bohmian mechanics the hidden variable
distributions from examples A and B overlap nevertheless with those
of example C and D in contradiction with the PBR result (the problem
is solved with the new axiomatic). }\end{figure}and identically
$P(4|1)=1/2$ of recording the photon in exit gate 4. Alternatively,
we can consider a  single photon wave packet coming from gate  2 and
at the end of the photon journey we will still get
$P(3|1)=P(4|1)=1/2$. From the point of view of the hidden variable
space we can write
\begin{eqnarray}
   P(4|1 \textrm{ or } 2)=\int P(3|\lambda)\rho(\lambda|\Psi_1 \textrm{ or } \Psi_2 )=1/2
\end{eqnarray} with `or' meaning exclusiveness.
Nothing can be said about the probabilities involved in the
integral. Now, if we consider superposed states such as
$|\pm\rangle=[|\Psi_1\rangle \pm i|\Psi_2\rangle]/\sqrt{2}$ the
photon will finish either in gate 3 or 4 with probabilities
$P(3|+)=P(4|-)=1$ and $P(4|+)=P(3|-)=0$. We here find us in the
orthogonal case of PBR theorem (i.e. $\langle
+|-\rangle=0$)\cite{Drezet3}. The deduction is thus straightforward
and we get  $\rho(\lambda|+ )\rho(\lambda|- )=0$  for all possible
$\lambda$ which means that the two densities of probability for
superposed states can not have any common intersecting support in
the
$\lambda$-space. This is what we should conclude if we consider a model accepting the PBR axiom $P(\alpha|\mathbf{a},\lambda)$.\\
However, this is not what happens in the pilot wave approach. In
this model where the spatial coordinates play a fundamental role we
don't have $\rho(\lambda|+/-)\rho(\lambda|\Psi_2 )=0$ neither we
have $\rho(\lambda|+/-)\rho(\lambda|\Psi_1)=0$ for every $\lambda$!
Indeed, half of the  relevant points of the wave packets $+$ or $-$
are common to $\Psi_1$ or $\Psi_2$. Actually, this is even worst
since we also have
$\rho(\lambda|+)\rho(\lambda|-)=\rho(\lambda|\pm)^2\neq 0$ for every
$\lambda$ in the full $\lambda$-support (sum of the two disjoint
supports associated with  $\Psi_1$ and $\Psi_2$). This is in
complete contradiction with PBR theorem `old' axiomatic (i.e., not
the version presented by me page 13). This is not surprising if we
remember that with pilot wave we have
$P(\alpha|\mathbf{a},\lambda,\Psi_0)$ and not
$P(\alpha|\mathbf{a},\lambda)$. For me what the PBR theorem shows is
that somehow those classical-like models obeying to the PBR
constraint $P(\alpha|\mathbf{a},\lambda)$ can not reproduce wave
particle duality: these models are therefore trivially useless. This
is actually not completely true because  we are here sticking too
much to the classical world with particle coordinates etc... If you
reject that classical framework you can still find some good models
reproducing the experiments and satisfying the PBR axiom
$P(\alpha|\mathbf{a},\lambda)$ but they don't look at all like
classical physics (see my proposals \cite{Drezet2}). However, if you
want to conserve some classical features like paths and positions
then you can use  Bohmian mechanics but you will now have
$P(\alpha|\mathbf{a},\lambda,\Psi_0)$ instead  of
$P(\alpha|\mathbf{a},\lambda)$! Furthermore, the PBR theorem is for
me very useful since if we accept to include explicitly the missing
axiom discussed before then we deduce with PBR that the kind of
`XIX$^{th}$ century like' epistemic model (i.e., imposing
$P(\alpha|\mathbf{a},\lambda)$ but contradicting Eq.~4) are necessarily condemned.\\
\indent What to conclude?  I reviewed  some of the fundamental
aspects of the pilot wave approach and I discussed the PBR theorem
within this context. Bohmian mechanics is for me the best available
ontology, but it will certainly one day be superseded by a better
theory justifying some of its magical assumptions. In this context
PBR's theorem, like Bell's one, is very useful for discussing the
pertinence of future and present hidden variable models. However,
this theorem should be formalized in order to discuss the best
existing models (like the one of de Broglie and Bohm) and therefore
equipped with a satisfying axiomatic. When this is done correctly
the difficult discussion concerning ontic and epistemic becomes
easier and the theorem strength is nicely enforced.\\
\underline{Post-scriptum:}\\
\indent I would like to briefly discuss a consequence of the PBR
theorem that M. Leifer~\cite{Leiferblog} called `the supercharged
EPR argument'. This argument is also discussed in a recent paper by
G.~Hetzroni and D.~Rohrlich \cite{Rohrlich} (focussing on the
relation between PBR and protective measurement; see also S.
Gao~\cite{Gao} on this topic). The argument runs as follows. Take an
EPR-like state i.e. a singlet state. This defines a pair of
entangled Q-bits. Now, if you project one of the two remote Q-bits
`Alice' along a basis (i.e., using a Bell procedure) the second
Q-bit `Bob'is projected in a specific state depending on the
outcomes obtained for Alice. However, if you admit the PBR theorem
but only consider, like Leifer did, the cases
`$P(\alpha|\mathbf{a},\lambda)$' (i.e. without the presence of
$\Psi_0$) then you could conclude the following:  The possible
states of Bob are depending on the basis choices for Alice. These
Bob states are different and in agreement with PBR these can not
overlap in the $\lambda-$space (see Eq.~4). However, the basis
choice for Alice can be done arbitrarily fast and therefore the Bob
state will be collapsed with arbitrary huge velocity into its
associated state. This would imply non-locality and this without
involving Bell theorem! This is very nice, but now we see the
interest of our new version of the PBR theorem: if we admit that the
conditional probabilities can depend on the quantum state $\Psi_0$
the deduction doesn't hold anymore because Eq.~4 is not true. Still,
the conclusion is perhaps correct because if
`$P(\alpha|\mathbf{a},\lambda)$' becomes
`$P(\alpha|\mathbf{a},\lambda,\Psi_0)$' we have apriori a clear non
local feature from the start (Bohmian mechanics is nonlocal after
all). In other words: projecting Bob in different states $\Psi_i$
means different dynamics `$P(\alpha|\mathbf{a},\lambda,\Psi_i)$'
which are enforced non locally by the projection of Alice outcomes.
It can be useful to be a bit more precise here.  By
`$P(\alpha|\mathbf{a},\lambda,\Psi_i)$' or
`$P(\alpha|\mathbf{a},\lambda)$' I mean the equivalent for the EPR
case of the notation used in this paper. But, of course since we
have two Q-bits and two sets of measurements characterized by -for
example- Stern and Gerlach directions $\mathbf{a}$ (for Alice) and
$\mathbf{b}$ (for Bob) we must precise a bit our notations. First,
$P(\beta,\alpha|\mathbf{b},\mathbf{a})$ means the probability for
finding the system with outcome $\alpha=\pm1$ for Alice if her
measurement device is aligned along $\mathbf{a}$ and $\beta=\pm1$
for Bob if his measurement device is aligned along $\mathbf{b}$. I
will omit the $\Psi_0$ notation for the singlet here since this is
the same state during all the reasoning. Then, using the $\lambda$
notations we will get with Bell
\begin{eqnarray}
P(\beta,\alpha|\mathbf{b},\mathbf{a})=\int
P(\beta,\alpha|\mathbf{b},\mathbf{a},\lambda)\rho(\lambda)d\lambda.
\end{eqnarray} We assume that $\rho(\lambda)$ is not depending on
$\mathbf{a}$, $\mathbf{b}$  because we don't like retro-causality
(those who don't agree could argue at that point) and we will
therefore accept this simple causal condition. Now, the
EPR-Leifer-PBR measurement is made in two steps: first, Alice is
projected and we get $\alpha$, then Bob and we get $\beta$. For this
reason we can instead of Eq.~9 write
\begin{eqnarray}
P(\beta,\alpha|\mathbf{b},\mathbf{a})=\int
P(\beta|\alpha,\mathbf{b},\mathbf{a},\lambda)dP(\alpha,\mathbf{b},\mathbf{a},\lambda)\nonumber\\
=\int
P(\beta|\alpha,\mathbf{b},\mathbf{a},\lambda)P(\alpha|\mathbf{b},\mathbf{a},\lambda)\rho(\lambda)d\lambda.
\end{eqnarray} Now, we have many probabilities.  The first one from the  right is
$\rho(\lambda)$ the density of probability in the initial hidden
variable space. The second is
$P(\alpha|\mathbf{b},\mathbf{a},\lambda)$ the conditional
probability for going from the initial state to a state where
Alice's outcome is projected to $\alpha$. This rigorously depends on
$\mathbf{a}$ and $\mathbf{b}$  but like for
$\rho(\mathbf{b},\mathbf{a},\lambda)=\rho(\lambda)$ this will be
simplified (using some causality prerequisites in this reference
frame) to
$P(\alpha|\mathbf{b},\mathbf{a},\lambda)=P(\alpha|\mathbf{a},\lambda)$
since the result of Alice can not depend on the not yet realized
outcome of Bob and device $\mathbf{b}$ if space like separation is
considered. Again, this is not a very general hypothesis (no
retro-causality) but I only accept it in order to stick to the
Bohmian framework. The last term is
$P(\beta|\alpha,\mathbf{b},\mathbf{a},\lambda)$ the conditional
probability to get $\beta$ for Bob knowing that we had $\alpha$ for
Alice and that we started from $\lambda$.  This is the PBR
probability discussed before. It depends on  the quantum state
$\Psi_i:=|\alpha\rangle$ associated with the possible outcomes for
Alice and depends also from the axes directions $\mathbf{a}$ and
$\mathbf{b}$. But wait, how do I know that
$P(\beta|\alpha,\mathbf{b},\mathbf{a},\lambda)$ should depend on
$\mathbf{a}$ and $\alpha$ ? No, problem guys: simply take Bell's
theorem with its non locality proof. From Eqs.~9, 10 and Bell we
have
$P(\beta,\alpha|\mathbf{b},\mathbf{a},\lambda)=P(\beta|\alpha,\mathbf{b},\mathbf{a},\lambda)P(\alpha|\mathbf{a},\lambda)\neq
P(\beta|\mathbf{b},\lambda)P(\alpha|\mathbf{a},\lambda)$ meaning
that $P(\beta|\alpha,\mathbf{b},\mathbf{a},\lambda)$ should depends
on $\textbf{a}$ and $\alpha$. A detailed calculation in the context
of Bohm theory would lead the same result. In other words accepting
the different causality axioms used here Bell theorem is necessary
anyway to get non locality. Few additional remarks are here
important. First, Leifer considered the case where the conditional
probabilities are not depending on the quantum state.  From our own
result this would imply that
$P(\beta|\alpha,\mathbf{b},\mathbf{a},\lambda)$ is independent from
$\textbf{a}$ and $\alpha$ in apparent contradiction with Bell!
However, this is not the case since it is not actually necessary to
remove the dependence on $\textbf{a}$: only $|\alpha\rangle$ should
be removed (in agreement with Leifer choice) so that Bell is safe
and indeed non-locality holds. Therefore, from this reasoning it is
difficult for me to see PBR as kind of proto-theorem  able to create
a `supercharged EPR argument' since Bell is with us all along. A
second remark concerns `wave function collapse' in the regime
involving Bohmian mechanics. Einstein, de Broglie, and Bohm didn't
like the wave function collapse: it looked as magic. Unless we
introduce a nonlinear process, like GRW did, this is not physical.
In the theory of de Broglie and Bohm there is no wave collapse.  The
different branches of the measuring process are all playing a role
even those with an `empty wave'. Still, in the effective this is the
same because the entanglement process between Alice and Bob breaks
the coherence between the different possible states of Bob if one do
a projective measurement on Alice. Every thing will be like if we
have a statistical mixture which is somehow equivalent to a collapse
since the quantum nature of the motion is now erased (in the sense
of a `which-path' experiment). Finally, I would like to point out
that non-locality is in the current Bohmian theory a very curious
thing. It clearly involves a kind of privileged reference frame or
`Aether' with a specific space-time foliation (see for
example~\cite{foliation}). This is not really covariant and we have
the feeling to return to the Lorentz-Poincar\'{e}'s time when the
relativity principle was clearly defined but when people tried to
save a privileged frame anyway. For me this again motivates
researches for a better theory.\\
\indent I would like to thank M.~Leifer, and the PBR authors for
very interesting discussions in 2012. I would like to thank the CNRS
for giving me the possibility to make at the same time experimental
/theoretical physics in such `fashionable topics' like
quantum-plasmonics \cite{Nanoletters} and letting me the opportunity
to do fundamental physics.

\end{document}